# CredSec: A Blockchain-based Secure Credential Management System for University Adoption

Md. Ahsan Habib
Dept. of CSE
KUET
Khulna, Bangladesh
Email: mahabib@cse.kuet.ac.bd

Md. Mostafijur Rahman
Dept. of CSE
DIU
Dhaka, Bangladesh
Email: mostafijur.cse@diu.ac

Nieb Hasan Neom
Software Engineer
Kona Software Lab Ltd
Dhaka, Bangladesh
Email: niebhasanneom0@gmail.com

*Abstract–* University education play a critical role in shaping intellectual and professional development of the individuals and contribute significantly to the advancement of knowledge and society. Generally, university authority has a direct control of students' result making and stores the credential in their local dedicated server. So, there is chance to alter the credential and also have a very high possibility to encounter various threats and different security attacks. To resolve these, we propose a blockchain-based secure credential management system (BCMS) for efficiently storing, managing and recovering credential without involving the university authority. The proposed BCMS incorporates a modified two-factor encryption (m2FE) technique, a combination of RSA cryptosystem and a DNA encoding to ensure credential privacy and an enhanced authentication scheme for teachers and students. Besides, to reduce size of the cipher credential and its conversion time, we use character to integer (C2I) table instead of ASCII table. Finally, the experimental result and analysis of the BCMS illustrate the effectiveness over state-of-the-art works.

## I. INTRODUCTION

Education is a fundamental catalyst for personal growth, societal well-being, and enhancement of global progress [1]. For this purpose, a university plays very significant role by providing quality education. However, it has substantial concern with this aspect and it is greatly affected by the assessment processes of the students. A university student is evaluated from various aspects including critical analysis of a problem, solving, evaluation, etc., which prepare them for the challenges of 21st century. However, the assessment credential is vital for both the student and university. Generally, the existing credential management system (CMS) of a university stores the credential to a local dedicated server (LDS). For further usage of the credential, the university authority retrieves the credential from the LDS [2]. Moreover, the LDS-based CMS may face various threats (i.e., single point failure, credential manipulation and/or revelation of the credential, etc.) and different security attacks (i.e., SQL injections, DDoS, cross-site scripting, etc.). Furthermore, any modification of the credential could devastate the student's career [3].

The activity of manipulating the credential is referred to as student credential tampering, which is not only unethical but also may life-threatening. Any individual with access on the credential may carry out this action. This behavior may also carries a risk of negative outcomes, such as falsifying a student's qualifications and accomplishments, causing erroneous college admissions or employment choices, and undermining the integrity of the educational system. For this reason, university may enact stringent security measures, such as secure storage systems and auditing processes, guarantee of the credential accuracy and prevention from credential tampering. However, a blockchain-based secure credential management system (BCMS) can efficiently handle these problems with high credential integrity and auditing [4].

In this study, by taking into account both the advantages of blockchain technology [5] and the existing CMS [6], we propose a BCMS to throttle the above mentioned threats and security problems. The significant contributions of this work are:

1. To develop a tamper-proof CMS that streamlines the operations of grade storage, management, and recovery without involving any third party. Any legit individual can securely access the credential.
2. To propose a modified two-factor encryption (m2FE) technique using an asymmetric cryptosystem and shuffled DNA encoding technique that ensures the confidentiality of the credential.
3. To design an enhanced authentication scheme that enables login, registration, and authentication process of a user.
4. To employ the C2I table that reduces the size of cipher credential and its conversion time (text to integer and vice versa).
5. To deploy both the local dedicated server (LDS) and blockchain server (BCS) that facilitate the tamper-proof nature along with credential verification.

## II. LITERATURE REVIEW

### A. LDS-based CMS

The proposed web-based CMS [7] makes it possible to efficiently enter student credential using the Microsoft Office Access as a database. However, the CMS has a number of drawbacks, including (*i*) the lack of a method to confirm the validity of certificates, (*ii*) the time needed for the system to validate forgeries, and (*iii*) the possibility of unauthorized users using the credential. Another web-based CMS [8] employs the ADO.NET but the CMS does not protect credential integrity, authenticity, and privacy, etc. However, malevolent attacks may change the credential, which could hinder students' academic progress and professional development.

To the best of our knowledge, almost all universities employ web-based CMS that depend on LDS, leaving them open to up to 20 security intrusions [9]. Besides, the attacker could attack the associated institution or organization in a various number of ways [7], [8]. Invalidated inputs, ineffective error management, unsecured direct object references, remote malicious file inclusion, confidentiality failure, etc. are some of the attacks. In addition, the LDS must also be concerned about single point failure. Here, the idea of blockchain technology can be used to solve these problems [10].

*B. BCS-based CMS*

The proposed BCMS in [11] uses blockchain technology to support student mobility. Specifically, the BCMS employs Ethereum public blockchain so student credential is used in a more flexible and user-oriented way. Hence, it is very convenient for the credit transfer approach when students move from one university to another university. Also, the student is able to see his/her result efficiently. However, here they did not focus on security issues.

The suggested BCMSs in [12] and [13] employ both the blockchain technology and IoT integration to protect the credential. The BCMS [12] uses permissioned Hyperledger Fabric with no encryption technique to encrypt the credential which could reveal the confidentiality. On the other hand, the BCMS [13] incorporates encryption technique to secure the credential but could face the credential manipulation by the authority.

Blockcrets [14], a BCMS that verifies certificates and other credentials, uses the Bitcoin blockchain to hold digital certificates. MIT, the University of Nicosia (UNIC), the University of Birmingham, etc. use it as universal verifier. Since Bitcoin is open source, the BCMS needs to include robust privacy safeguards. Besides, the transaction fee associated with Bitcoin is much higher than those of Ethereum and other blockchains [15]. However, Digicert [16] overcome this limitation by employing the Ethereum, but the BCMS requires more efficient access to the credentials. Furthermore, there exist several BCMSs like Recordkeeper [17], Smartcert [18], etc., but they have security weaknesses, like [17] has the integrity issue in case of the credential usage by any third party and [18] needs to improve the credential privacy along with the ease access of legal users.

Concerning the university student, the BCMS in [11] employs asymmetric cryptosystem to facilitate the privacy of credential having a recovery policy of the credential. The BCMS uses both LDS and BCS while the Ethereum as BCS. However, the BCMS deploys a centralized university authority who has a full control over the credential. Hence, there is a high chance of manipulation and/or abuse the student credential. Besides, the BCMS does not mention any authentication mechanism for identifying students and teachers.

*C. Data confidentiality with two-factor encryption*

The study [19] protects data using two-factor encryption (2FE) by combining an asymmetric cryptosystem and dynamic DNA encoding. It employs RSA, ElGamal, and Paillier cryptosystem as asymmetric one and then encodes with DNA bases. Before encoding, it injects dummy numbers into the actual data using the Fibonacci series. However, it is acceptable with small amount of data but produces high overhead for large data as Fibonacci series produce very big number after a certain point. Another work in [20] also uses the Fibonacci series is equally impractical for high overhead.

After analyzing the above-mentioned works, we develop a BCS-based CMS for efficiently storing, maintaining, and retrieving credential by overcoming the aforementioned limitations. We ensure high privacy of the credential by employing modified two-factor encryption (m2FE) and store on both the LDS and BCS. Besides, we design an enhanced authentication scheme to enable login, registration, and authentication process of a user. In addition, by employing the C2I table we reduce the cipher credential size along with its conversion time.

### III. PRELIMINARIES

*A. C2I table*

The C2I table [21] is a character to integer mapping approach that is efficient compared to ASCII table. To convert a single character into its corresponding integer value, the ASCII table requires 3 decimal digits whereas the C2I table requires only 2 decimal digits as shown in Table I. In the ASCII table there exist 256 individual characters but most of them remain useless (most of the cases). On the other hand, the C2I table considers only 95 individual useful characters and as a result it is needed only 2 decimal digits when converting a single character to its corresponding integer value. Mathematically, it clearly shows that, by employing the C2I table in replace of the ASCII table in the conversion of text to integer data, around 33% overhead (size and time) will be less. We employ the C2I table in the proposed CMS to reduce the overhead.

TABLE I. CHARACTER TO INTEGER (C2I) MAPPING

| Ch | Int | Ch | Int | Ch | Int | Ch | Int | Ch | Int | Ch | Int |
|---|---|---|---|---|---|---|---|---|---|---|---|
| 0 | 01 | H | 18 | Y | 35 | p | 52 | ‘ | 69 | ^ | 86 |
| 1 | 02 | I | 19 | Z | 36 | q | 53 | ( | 70 | _ | 87 |
| 2 | 03 | J | 20 | a | 37 | r | 54 | ) | 71 | [ | 88 |
| 3 | 04 | K | 21 | b | 38 | s | 55 | * | 72 | \ | 89 |
| 4 | 05 | L | 22 | c | 39 | t | 56 | + | 73 | ] | 90 |
| 5 | 06 | M | 23 | d | 40 | u | 57 | , | 74 | ` | 91 |
| 6 | 07 | N | 24 | e | 41 | v | 58 | - | 75 | ~ | 92 |
| 7 | 08 | O | 25 | f | 42 | w | 59 | . | 76 | { | 93 |
| 8 | 09 | P | 26 | g | 43 | x | 60 | / | 77 | \| | 94 |
| 9 | 10 | Q | 27 | h | 44 | y | 61 | : | 78 | } | 95 |
| A | 11 | R | 28 | i | 45 | z | 62 | ; | 79 | | |
| B | 12 | S | 29 | j | 46 | | 63 | < | 80 | | |
| C | 13 | T | 30 | k | 47 | ! | 64 | = | 81 | | |
| D | 14 | U | 31 | l | 48 | " | 65 | > | 82 | | |
| E | 15 | V | 32 | m | 49 | # | 66 | ? | 83 | | |
| F | 16 | W | 33 | n | 50 | % | 67 | @ | 84 | | |
| G | 17 | X | 34 | o | 51 | & | 68 | $ | 85 | | |

*B. Blockchain*

Blockchain is a distributed, decentralized ledger and tamper-proof information storage technology that confirms data integrity. It is used for store historical record of all transactions that have taken place across a peer-to-peer network. Nowadays in many different types of applications it is highly used, e.g., financial services, healthcare, supply chain management, public administration, etc. In blockchain the data are stored linearly in blocks and cryptographically linked together to form a chain. Every block contains a timestamp, a cryptographic hash pointer of the previous block and transaction of data. The immutability property of the blockchain is obtained through the hash pointer that creates the link with previous block.

*C. m2FE*

Based on the deployed two-factor encryption (2FE) in [22], here we develop a modified 2FE (m2FE) that comprises of a public key encryption and a DNA encoding method. Here, the plaintext is encrypted by using an asymmetric cryptosystem, i.e., RSA encryption method and then DNA encoding technique is employed to enrich the data privacy. As m2FE consists of two encryption mechanism, i.e., m2FE = {RSA, DNA}, at first, we present the formal definition of RSA as RSA = {Setup$_r$, KeyGen$_r$, Enc$_r$, Dec$_r$} and they are as follows.

Setup$_r$ ($1^\lambda$) → ($p, q, N$): The probabilistic parameter setup algorithm takes a security parameter $\lambda \in \mathbb{N}$ as input, and outputs two prime number $p, q$ and $N$, where $N = p \times q$.

KeyGen$_r$ ($p, q, N$) → ($e, d$): The probabilistic key generation algorithm takes two prime number $p$ and $q$ as input, and outputs public key $e$ and private key $d$.

Enc$_r$ ($m, e, N$) → $c$: The probabilistic encryption algorithm takes plaintext $m$, public key $e$ and $N$ as input, and outputs ciphertext $c = m^e \mod N$.

Dec$_r$ ($c, d, N$) → $m$: The deterministic decryption algorithm takes ciphertext $c$, private key $d$ and $N$ as input, and outputs $m = c^d \mod N$.

Now, we present the formal definition of DNA as DNA = {Setup$_d$, KeyGen$_d$, DumGen$_d$, Enc$_d$, Dec$_d$} and they are as follows.

Setup$_d$ ($1^K$) → ($S, T$): The probabilistic parameter setup algorithm takes a security parameter $K \in \mathbb{N}$ as input, and outputs two number $S$ and $T$.

KeyGen$_d$ ($S, T$) → $DK$: The probabilistic key generation algorithm takes two number $S$ and $T$ as input, and outputs a Ɛ-bit DNA key $DK$ = toBinary($ln(S) \times T^2 + ln(T) \times S^2$).

DumGen$_d$ ($C_{i-1}, C_i, S, T$) → $\Delta$-digit integer Ґ: The probabilistic dummy number generation algorithm takes $C_{i-1}$, $C_i, S, T$ as input, and outputs a $\Delta$-digit integer Ґ. In details, it first generates α, β, Λ, Δ, and Ψ where α = $len(C_{i-1})$, β = $len(C_i)$, Λ = sqrt (($\alpha + \beta)^2 / \alpha \times \beta$), Δ = Λ×$S \mod T$, and Ψ = 1 + ((Λ+$S \mod (S-T)) \mod 2$). The chunk ($C_{i-1}$ or $C_i$) from where Ґ gets selected is determined by the Λ i.e., if Λ is odd, it selects $C_i$, otherwise $C_{i+1}$. And the Ψ determines the portion of the chunk, from where dummy bits are picked i.e., if Ψ is 1, it selects the first Δ-digit, otherwise the last Δ-digit.

Enc$_d$ ($m, DK, \Upsilon$) → $c$: The probabilistic encryption algorithm takes binary plaintext $m$ and a rule $\Upsilon$, and outputs ciphertext $c = \sum_{i=0}^{\lfloor len(z)/2 \rfloor} \text{DNA}((z[2i+1], z[2i+2]), \Upsilon)$ where z = $m \oplus DK$. As DNA has 4 bases (A, T, C, G), it can have 4! different mapping schemes from where a rule $\Upsilon$ get picked.

Dec$_d$ ($c, DK, \Upsilon$) → $m$: The deterministic decryption algorithm takes ciphertext $c$ and a rule $\Upsilon$ as input, and outputs binary $m = \sum_{i=0}^{\lfloor len(m)/2 \rfloor} \text{BIN}(x[i], \Upsilon)$, where $x = c \oplus DK$.

As the m2FE is combination of RSA and DNA, we can summarize the m2FE encryption and decryption processes in the following algorithm 1 and algorithm 2, respectively.

However, before m2FE decryption process, the following credential verification algorithm regarding the integrity checking of the ciphertext is done.

Ver ($H_c'$, $H_c$) → $b$: The deterministic credential verification algorithm takes the calculated hash $H_c'$ and retrieved hash $H_c$ as input, and outputs $b$ where $b \in \{0,1\}$. If $b$ is 1 i.e., $H_c'$ = $H_c$ then the integrity of $c$ is maintained otherwise the following credential recovery algorithm is executed. Note that, the SHA256 algorithm is used for hashing purpose.

Rec ({STD$_{INFO}$}) → $c$: The deterministic credential recovery algorithm takes the information of a student STD$_{INFO}$ as input, obtains cipher credential $c$ from the BCS, updates the LDS and outputs $c$ which is further sent to the STD.

| *Algorithm 1:* m2FE encryption process |
|---|
| 1: **Input:** $m, e, S, T, \Upsilon$ |
| 2: **Output:** $c$ |
| 3: $A$ ← C2I ($m$) |
| 4: $E$ [1…$n$] ← splitToChunk ($A$) |
| 5: **For each** $ch$ in $E$ **do:** |
| 6:   $C_i$ ← Enc$_r$ ($ch, e, N$) |
| 7:   $P$ ← DumGen$_d$ ($C_{i-1}, C_i, S, T$) |
| 8:   $Q$.append ($P$) |
| 9: **End for** |
| 10: $C_G$ ← makeSingleString ($Q$ [1…$n$]) |
| 11: $B_G$ ← convertToBinary ($C_G$) |
| 12: $c$ ← Enc$_d$ ($B_G, DK, \Upsilon$) |
| 13: **Return** $c$ |

| *Algorithm 2:* m2FE decryption process |
|---|
| 1. **Input:** $c, d, S, T, \Upsilon$ |
| 2. **Output:** $m$ |
| 3. $B_G$ ← Dec$_d$ ($c, DK, \Upsilon$) |
| 4. $C_G$ ← convertToInteger ($B_G$) |
| 5. $Q$ [1…$n$] ← splitToChunk ($C_G$) |
| 6. **For each** $C_i$ in $Q$ **do:** |
| 7.   $P$ ← discardDumGen$_d$ ($C_{i-1}, C_i, S, T$) |
| 8.   $ch$ ← Dec$_r$ ($P, d, N$) |
| 9.   $E$.append ($P$) |
| 10. **End for** |
| 11. $A$ ← makeSingleString ($E$ [1…$n$]) |
| 12. $m$ ← C2I ($A$) |
| 13. **Return** $m$ |

IV. PROPOSED BCMS

In this section, we present the system model and formal definition of the proposed BCMS, including roles of each entity in BCMS, syntax of BCMS, workflow of BCMS, threat model and security model.

*A. Roles of each Entity*

The proposed BCMS consists of mainly six entities: certification authority (CTA), instructor (INS), credential merging system (CMS), student (STD), blockchain server (BCS), and local dedicated server (LDS).

**CTA** The CTA is responsible to verify the identities of STD and INS. It binds cryptographic keys to the STD whereas provides nonce to the INS.

**INS** The INS first creates the grade of a STD according to the course he has taken. Then the INS submits the grade to the CMS.

**CMS** The CMS facilitates two main services, namely, (*i*) it merges student's grades and calculates final credential, (*ii*) it provides the credential when a legit STD request it.

**STD** The STD wants to keep his/her credential secure and access the credential by using the private key.

**BCS** The BCS is responsible for storing both the credential and its hash value of each individual STD.

**LDS** The LDS is also responsible for storing both the credential and the hash value of each individual STD.

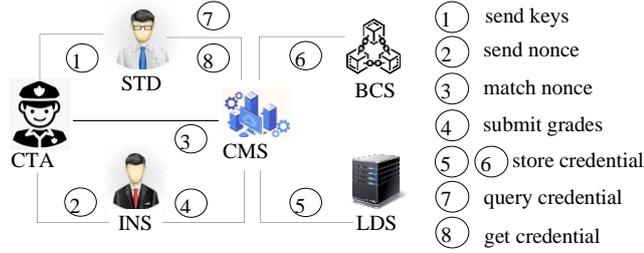

Fig. 1: System Model of the proposed BCMS

*B. Syntax of BCMS*

A blockchain-based secure credential management system consist of a tuple of algorithms as follows. Note that, a few of the following algorithms outputs $b \in \{0,1\}$ where the value 1 and 0 respectively represent the successful and unsuccessful operation.

INSReg$_{CTA}$ (*email*, *ID*) → Ñ: This INS registration algorithm is run by the CTA that takes verified *email* and unique identification *ID* of INS as input, and outputs a nonce Ñ.

INSReg$_{CMS}$ (*email*, *password*, *ID*, Ñ) → *b*: This INS registration algorithm is run by the CMS that takes *email*, *password*, *ID*, and nonce Ñ of INS as input, and outputs *b* where $b \in \{0,1\}$.

STDReg$_{CTA}$ (*email*, *roll*) → (*e*, *d*, *DK*): This STD registration algorithm is run by the CTA that takes verified *email* and *roll* of STD as input, and sends (*e*, *d*, *DK*) after running the following m2FE.KeyGen (*p*, *q*, *N*, *S*, *T*).

STDReg$_{CMS}$ (*email*, *password*, *roll*) → *b*: This STD registration algorithm is run by the CMS that takes *email*, *password*, and *roll* of STD as input, and outputs *b* where $b \in \{0,1\}$.

INSLog$_{CMS}$ (*email*, *password*, Ñ) → *b*: This INS login algorithm is run by the CMS that takes *email*, *password*, and Ñ of INS as input, and outputs *b* where $b \in \{0,1\}$.

STDLog$_{CMS}$ (*email*, *password*) → *b*: This STD login algorithm is run by the CMS that takes *email* and *password* of STD as input, and outputs *b* where $b \in \{0,1\}$.

m2FE.Setup ($1^\lambda$) → (*p*, *q*, *N*, *S*, *T*): The setup algorithm is run by the CTA that takes a security parameter $\lambda \in \mathbb{N}$ as input, and outputs *p*, *q*, *N*, *S*, and *T*.

m2FE.KeyGen (*p*, *q*, *N*, *S*, *T*) → (*e*, *d*, *DK*): The key generation algorithm is run by the CTA which takes the *p*, *q*, *N*, *S*, and *T* as input, and outputs public key *e*, private key *d*, and DNA key *DK*.

m2FE.DumGen ($C_{i-1}$, $C_i$, *S*, *T*) → ℂ: The dummy integer generation algorithm is run by the INS that takes the $C_i$, $C_{i+1}$, *S*, and *T* as input, and outputs ℂ which is a Δ-digit integer.

m2FE.Enc (*m*, *e*, *N*, *DK*, ϒ) → *c*: The encryption algorithm is run by the INS which takes the plain credential *m*, public key *e*, *N*, DNA key *DK*, and a rule ϒ as input, and outputs cipher credential *c* as depicted in Algorithm 1. Note that, this algorithm requires the dummy integer generation algorithm m2FE.DumGen ($C_{i-1}$, $C_i$, *S*, *T*) → ℂ in between every two chunks.

m2FE.Ver ($H_c'$, $H_c$) → *b*: The credential verification algorithm is run by the STD which takes the calculated hash $H_c'$ and retrieved hash $H_c$ as input, and outputs *b* where $b \in \{0,1\}$.

m2FE.Rec (*email*, *password*, *roll*) → *c*: The credential recovery algorithm is run by the CMS that takes the *email*, *password*, and *roll* of a STD and obtains cipher credential *c* from the BCS, updates the LDS and outputs *c* which is further sent to the STD.

m2FE.Dec (*c*, *d*, *N*, *DK*, ϒ) → *m*: The decryption algorithm is run by the STD which takes cipher credential *c*, private key *d*, *N*, DNA key *DK*, and a rule ϒ as input, and outputs plain credential *m* as depicted in Algorithm 2.

*C. The workflow of BCMS*

According to the system model in Fig. 1, the workflow of our proposed BCMS includes four phases: entity registration, secure key distribution, credential uploading, and credential retrieving.

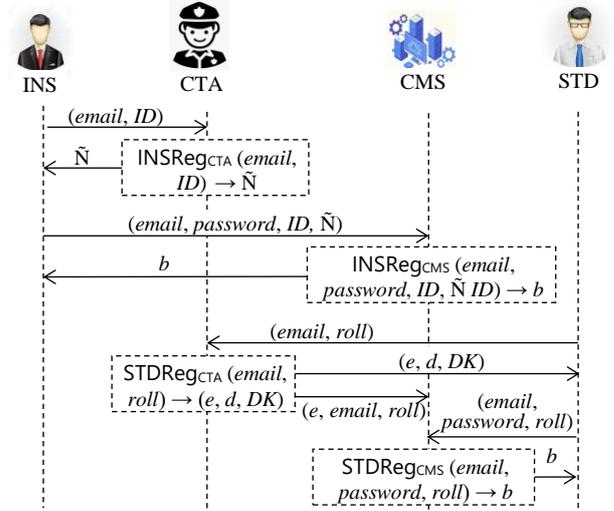

Fig. 2. Entity Registration in BCMS

***Entity registration***: Each entity namely INS and STD needs to get registered first, otherwise they don't interact within BCMS. The registration phase is depicted as follows and Fig. 2 illustrates the workflow.

*INS registration*: The CTA runs the INS registration algorithm INSReg$_{CTA}$ (*email*, *ID*) → Ñ to get a nonce Ñ. Both CTA and INS keep the Ñ secret. Then the CMS runs the INS registration algorithm INSReg$_{CMS}$ (*email*, *password*, *ID*, Ñ) → *b* that outputs *b* = 1 for successful registration.

*STD registration*: The CTA runs the STD registration algorithm STDReg$_{CTA}$ (*email*, *roll*) → (*e*, *d*, *DK*) to get the public-private key pair (*e*, *d*) and DNA key *DK*. Then the

CMS runs the STD registration algorithm STDReg$_{CMS}$ (*email*, *password*, *roll*) → *b* that outputs *b* = 1 for successful registration. Besides, the CTA sends (*e*, *email*, *roll*) to the CMS. Upon receiving, the CMS keeps the (*e*, *email*, *roll*) to itself.

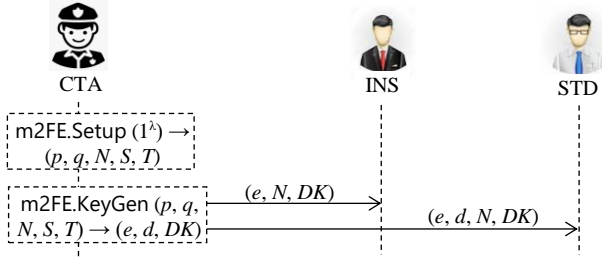

Fig. 3. Secure Key Distribution in BCMS

***Secure key distribution***: Fig. 3 illustrates the workflow of secure key distribution where the INS and STD get their corresponding encryption and/or decryption keys. The secure key distribution phase is described as follows.

At first, the CTA runs m2FE.Setup ($1^\lambda$) → (*p*, *q*, *N*, *S*, *T*) to get the *p*, *q*, *N*, *S*, and *T* and then runs m2FE.KeyGen (*p*, *q*, *N*, *S*, *T*) → (*e*, *d*, *DK*) to achieve the public-private key pair (*e*, *d*) and DNA key *DK*. After that, the CTA sends (*e*, *N*, *DK*) and (*e*, *d*, *N*, *DK*) to the INS and STD, respectively.

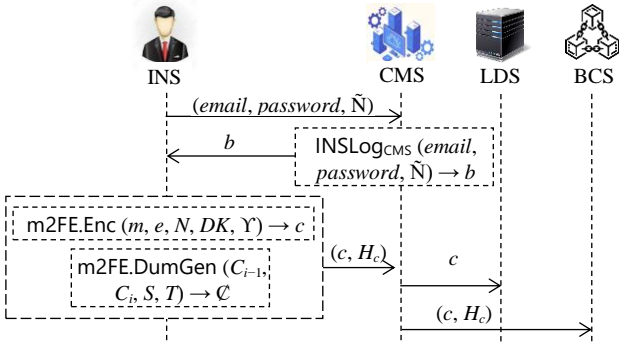

Fig. 4. Credential Uploading in BCMS

***Credential uploading***: Fig. 4 shows the credential uploading phase in the proposed BCMS. This phase consists of the following major sub-phases: *INS login*, *credential encryption*, and *credential storing*.

*INS login*: The CMS runs the login algorithm INSLog$_{CMS}$ (*email*, *password*, Ñ) → *b* that outputs 1 to INS to get entered into the CMS.

*Credential encryption*: The INS runs the credential encryption algorithm m2FE.Enc (*m*, *e*, *N*, *DK*, ϒ) → *c* that requires the dummy integer generation algorithm m2FE.DumGen ($C_{i−1}$, $C_i$, *S*, *T*) → ℂ in between every two chunks to increase the level of confusion. After encryption operation, the INS submits the cipher credential *c* of the corresponding STD to the CMS along with the hash of the cipher credential $H_c$.

*Credential storing*: Upon receiving the (*c*, $H_c$), first the CMS sends the *c* to the LDS and then (*c*, $H_c$) to the BCS for storage purpose.

***Credential retrieval***: Fig. 5 shows the credential retrieval phase in the proposed BCMS. This phase consists of the following major sub-phases: *STD login*, *credential acquisition*, *credential verification and recovery*, and *credential decryption*.

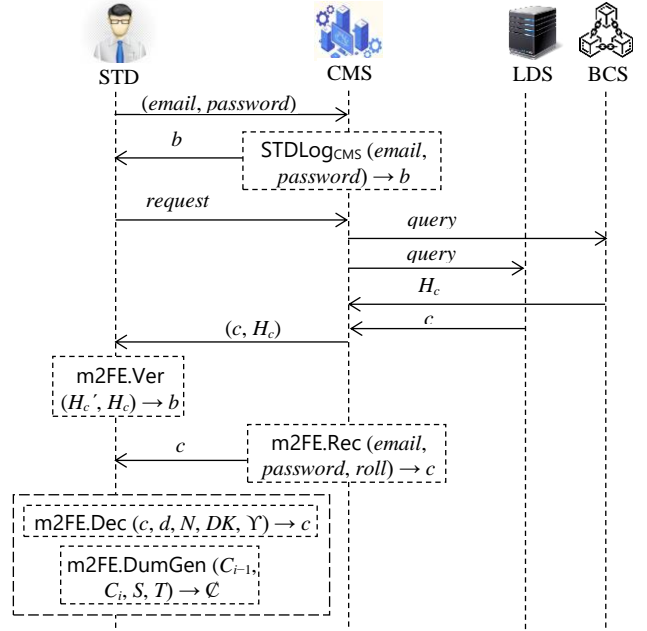

Fig. 5. Credential Retrieval in BCMS

*STD login*: The CMS runs the login algorithm STDLog$_{CMS}$ (*email*, *password*) → *b* that outputs 1 to STD to get entered into the CMS.

*Credential acquisition*: The STD requests his/her credential from CMS and CMS sends back the (*c*, $H_c$) after acquiring *c* from the LDS and $H_c$ from BCS.

*Credential verification and recovery*: The STD runs the credential verification algorithm m2FE.Ver ($H_c'$, $H_c$) → *b* to verify the integrity of *c*. If *b* is 0 (i.e., $H_c' \neq H_c$ which means that the integrity of *c* is breached in the LDS as the BCS is assumed to be immutable) then STD informs it to CMS and the CMS runs the credential recovery algorithm m2FE.Rec (*email*, *password*, *roll*) → *c* to update the LDS and sends back *c* to the STD.

*Credential decryption*: Finally, the STD runs the credential decryption algorithm m2FE.Dec (*c*, *d*, *N*, *DK*, ϒ) → *c* that requires to discard the dummy integer generated from the dummy integer generation algorithm m2FE.DumGen ($C_{i−1}$, $C_i$, *S*, *T*) → ℂ.

### D. Threat model

The CTA is a trusted entity, who genuinely completes the entity registration phase and issues the keys for different operations.

The INS and CMS are semi-trusted entities. The INS honestly performs the STD's grade generation, encryption and submission to the CMS. However, INS may try to learn any information without authorization. The CMS honestly stores, acquires and sends the credential of a STD, while it may try to learn about the credential of the STD.

The STD is an untrusted entity, who may try to modify his/her own credential and try to learn other STD's credential.

Here, we define three security models, i.e., indistinguishability model (IND-MOD), unforgeability model (UNF-MOD), and immutability model (IMU-MOD), to formalize the attacks within the threat model, considering the semi-trusted INS and CMS as well as the untrusted STD. Specifically:

**IND-MOD** This model preserves the confidentiality of credential, permitting only an authorized student with a valid decryption key to unveil the credential.

**UNF-MOD** This model guarantees the uprightness of credential, thwarting an unauthorized student from forging the credential without a valid decryption key.

**IMU-MOD** This model ensures the tamper-proofing, allowing the modification of the credential only by the CMS upon requesting.

*E. Security model*

### 1. IND-MOD

**Definition.** The security definition of the indistinguishability model in the proposed BCMS is based on the following game between a challenger $\mathcal{C}$ and an adversary $\mathcal{A}$.

**Initialization.** By taking a security parameter $\lambda \in \mathbb{N}$, $\mathcal{C}$ runs the setup algorithm m2FE.Setup $(1^\lambda) \rightarrow (p, q, N, S, T)$ and sends $p, q, N, S$, and $T$ to $\mathcal{A}$.

**Query.** $\mathcal{A}$ is allowed to make following oracle queries.

- $O_{\text{m2FE.KeyGen}}(p, q, N, S, T)$: By taking the $p, q, N, S$, and $T$ as input, $\mathcal{C}$ runs the key generation algorithm m2FE.KeyGen $(p, q, N, S, T) \rightarrow (e, d, DK)$ and sends back public key $e$, private key $d$, and DNA key $DK$ to $\mathcal{A}$.

- $O_{\text{m2FE.DumGen}}(C_{i-1}, C_i, S, T)$: By taking the $C_{i-1}, C_i, S$, and $T$ as input, $\mathcal{C}$ runs the dummy integer generation algorithm m2FE.DumGen $(C_{i-1}, C_i, S, T) \rightarrow \mathcal{C}$ and generates $\Delta$-digit integer $\mathcal{C}$.

- $O_{\text{m2FE.Enc}}(m_0, e, N, DK, \Upsilon)$: By taking the plain credential $m$, public key $e$, $N$, DNA key $DK$, and a rule $\Upsilon$ as input, $\mathcal{C}$ runs the encryption algorithm m2FE.Enc $(m_0, e, N, DK, \Upsilon) \rightarrow c_0$ (that also requires the dummy integer generation algorithm m2FE.DumGen $(C_{i-1}, C_i, S, T) \rightarrow \mathcal{C}$) and sends back cipher credential $c$ to $\mathcal{A}$.

**Challenge.** $\mathcal{A}$ submits plain credential $(m_0, m_1)$ as well as $(e, N, DK, \Upsilon)$. $\mathcal{C}$ picks $t \in \{0, 1\}$ randomly and generates the corresponding cipher credential $c_t$ which is sent back to $\mathcal{A}$.

**Guess.** $\mathcal{A}$ guesses a random bit $r \in \{0, 1\}$. $\mathcal{A}$ wins if $t = r$. We can say that the developed BCMS is indistinguishability based secure if the following gain of a PPT adversary $\mathcal{A}$ is negligible.

$^{\text{IND-MOD}}\text{Gain}_{\mathcal{A}}(1^\lambda) = |\Pr[t = r] - \tfrac{1}{2}| \approx \text{negl}(\lambda)$

### 2. UNF-MOD

**Definition.** The security definition of the unforgeability model in the proposed BCMS is based on the following game between a challenger $\mathcal{C}$ and an adversary $\mathcal{A}$.

**Initialization.** By taking a security parameter $\lambda \in \mathbb{N}$, $\mathcal{C}$ runs the setup algorithm m2FE.Setup $(1^\lambda) \rightarrow (p, q, N, S, T)$ and sends $p, q, N, S$, and $T$ to $\mathcal{A}$.

**Query.** $\mathcal{A}$ is allowed to make following oracle queries.

- $O_{\text{m2FE.KeyGen}}(p, q, N, S, T)$: Similar oracle of IND-MOD.
- $O_{\text{m2FE.DumGen}}(C_{i-1}, C_i, S, T)$: Similar oracle of IND-MOD.
- $O_{\text{m2FE.Dec}}(c_0, d, N, DK, \Upsilon)$: By taking the cipher credential $c_0$, private key $d$, $N$, DNA key $DK$, and a rule $\Upsilon$ as input, $\mathcal{C}$ runs the decryption algorithm m2FE.Dec $(c_0, d, N, DK, \Upsilon) \rightarrow m_0$ (that also requires the dummy integer generation algorithm m2FE.DumGen $(C_{i-1}, C_i, S, T) \rightarrow \mathcal{C}$) and sends back plain credential $m_0$ to $\mathcal{A}$.

**Challenge.** $\mathcal{A}$ submits cipher credential $(c_0, c_1)$ and $\mathcal{C}$ picks $t \in \{0, 1\}$ randomly and generates the corresponding plain credential $m_t$ which is sent back to $\mathcal{A}$.

**Guess.** $\mathcal{A}$ guesses a random bit $r \in \{0, 1\}$. $\mathcal{A}$ wins if $t = r$. We can say that the developed BCMS is unforgeability based secure if the following gain of a PPT adversary $\mathcal{A}$ is negligible.

$^{\text{UNF-MOD}}\text{Gain}_{\mathcal{A}}(1^\lambda) = |\Pr[t = r] - \tfrac{1}{2}| \approx \text{negl}(\lambda)$

### 3. IMU-MOD

**Definition.** The security definition of the immutability model in the proposed BCMS is based on the following game between a challenger $\mathcal{C}$ and an adversary $\mathcal{A}$.

**Initialization 1.** $\mathcal{A}$ updates the LDS with the forged $c_0{'}$ and try to convince $\mathcal{C}$ that $c_0 = c_0{'}$.

**Query 1.** $\mathcal{A}$ is allowed to make following oracle queries.

- $O_{\text{m2FE.KeyGen}}(p, q, N, S, T)$: Similar oracle of IND-MOD.
- $O_{\text{m2FE.DumGen}}(C_{i-1}, C_i, S, T)$: Similar oracle of IND-MOD.
- $O_{\text{m2FE.Enc}}(m_0, e, N, DK, \Upsilon)$: Similar oracle of UNF-MOD.
- $O_{\text{m2FE.Dec}}(c_0, d, N, DK, \Upsilon)$: Similar oracle of UNF-MOD.
- $O_{\text{m2FE.Ver}}(H_{c0}{'}, H_{c0})$: By taking the calculated hash $H_{c0}{'}$ and retrieved hash $H_{c0}$, $\mathcal{C}$ runs the credential verification algorithm m2FE.Ver $(H_{c0}{'}, H_{c0}) \rightarrow b$, verify the retrieved cipher credential $c_0$ and sends the update to $\mathcal{A}$.
- $O_{\text{m2FE.Rec}}(email, password, roll)$: By taking the $email$, $password$, and $roll$ of a STD, $\mathcal{C}$ runs the credential recovery algorithm m2FE.Rec $(email, password, roll) \rightarrow c_0$ and updates the LDS with $c_0$ obtains from the BCS.

**Challenge 1.** $\mathcal{A}$ updates the LDS with a forged credential $c_0{'}$ and try to deduce $H_{c0}$ (i.e., similar hash of $c_0$) using hash collision property. $\mathcal{C}$ verifies it and sends the update to $\mathcal{A}$.

**Guess 1.** $\mathcal{A}$ guesses a hash $\Pi$. $\mathcal{A}$ wins if $H_{c0} = \Pi$. We can say that the developed BCMS is immutability based secure if the following gain of a PPT adversary $\mathcal{A}$ is negligible.

$^{\text{IMU-MOD}}{}_1\text{Gain}_{\mathcal{A}}(1^\lambda) = \Pr[H_{c0} = \Pi] \approx \text{negl}(\lambda)$

**Initialization 2.** $\mathcal{A}$ updates the $n$th block of BCS having block hash $BH_{cur}$ with the forged $c_0{'}$ (hence new block hash is $BH_{new}$) and try to convince $\mathcal{C}$ that $BH_{new} = BH_{cur}$.

**Query 2.** This part is similar to Query 1.

**Challenge 2.** $\mathcal{A}$ updates the $n$th block having current block hash $BH_{cur}$ of BCS with a forged credential $c_0'$ and its hash $H_{c0}'$. $\mathcal{A}$ try to deduce the new block hash ($BH_{new}$) similar to $BH_{cur}$ using hash collision property.

**Guess 2.** $\mathcal{A}$ guesses a hash $\Gamma$. $\mathcal{A}$ wins if $BH_{cur} = \Gamma$. We can say that the developed BCMS is immutability based secure if the following gain of a PPT adversary $\mathcal{A}$ is negligible.

$$^{\text{IMU-MOD}}{}_2\text{Gain}_{\mathcal{A}}(1^{\lambda}) = \Pr[BH_{cur} = \Gamma] \approx \text{negl}(\lambda)$$

## V. SECURITY PROOFS

**Theorem 1.** *If the developed modified two-factor encryption* m2FE *is* IND-MOD-CPA*, then $\mathcal{A}$ breaks indistinguishable based security of the proposed BCMS with a negligible probability.*

**Proof.**

**Assumption.** Suppose $\mathcal{A}$ can break the indistinguishability security of the proposed BCMS with a non-negligible advantage.

**Simulator.** We can build a simulator $\mathcal{B}$ to break the security of the underlying m2FE run by $\mathtt{S}$. $\mathcal{B}$ randomly picks a security parameter $\lambda' \in \{1, \ldots, \Omega\}$ and provides various oracle queries (e.g., $O_{\text{m2FE.KeyGen}}$, $O_{\text{m2FE.DumGen}}$, $O_{\text{m2FE.Enc}}$, etc.) to $\mathcal{A}$ based on the $\lambda'$, likewise in section IV-*E*.

**Challenge.** $\mathcal{A}$ submits two plain credentials ($m_0$, $m_1$). $\mathcal{B}$ picks $t \in \{0, 1\}$ randomly and generates the corresponding cipher credential $c_t$ which is sent back to $\mathcal{A}$ and $\mathtt{S}$.

**Guess.** $\mathcal{A}$ guesses a random bit $r \in \{0, 1\}$ and sends $r$ to $\mathcal{B}$. Then $\mathcal{B}$ forwards it to $\mathtt{S}$. $\mathcal{A}$ wins if $\mathtt{S}$ finds $t = r$. Here, $\mathcal{A}$ is able to break the IND-MOD of the proposed BCMS only when it is able break the IND-MOD of the $\mathcal{B}$ (i.e., m2FE). However, the developed m2FE has guessing gain $1/\Omega$, i.e., $^{\text{IND-MOD}}\text{Gain}_{\mathcal{A}}(1^{\lambda}) = 1/\Omega$. Hence, we can say that the developed BCMS is indistinguishability based secure as the $^{\text{IND-MOD}}\text{Gain}_{\mathcal{A}}(1^{\lambda})$ of $\mathcal{A}$ is negl ($\lambda$).

**Theorem 2.** *If the developed modified two-factor encryption* m2FE *is* UNF-MOD-COA*, then $\mathcal{A}$ breaks unforgeable based security of the proposed BCMS with a negligible probability.*

**Proof.**

**Assumption.** Suppose $\mathcal{A}$ can break the unforgeability security of the proposed BCMS with a non-negligible advantage.

**Simulator.** We can build a simulator $\mathcal{B}$ to break the security of the underlying m2FE run by $\mathtt{S}$. $\mathcal{B}$ randomly picks a security parameter $\lambda' \in \{1, \ldots, \Omega\}$ and provides various oracle queries (e.g., $O_{\text{m2FE.KeyGen}}$, $O_{\text{m2FE.DumGen}}$, $O_{\text{m2FE.Dec}}$, etc.) to $\mathcal{A}$ based on the $\lambda'$, likewise in section IV-*E*.

**Challenge.** $\mathcal{A}$ submits two cipher credentials ($c_0$, $c_1$). $\mathcal{B}$ picks $t \in \{0, 1\}$ randomly and generates the corresponding plain credential $c_t$ which is sent back to $\mathcal{A}$ and $\mathtt{S}$.

**Guess.** $\mathcal{A}$ guesses a random bit $r \in \{0, 1\}$ and sends $r$ to $\mathcal{B}$. Then $\mathcal{B}$ forwards it to $\mathtt{S}$. $\mathcal{A}$ wins if $\mathtt{S}$ finds $t = r$. Here, $\mathcal{A}$ is able to break the UNF-MOD of the proposed BCMS only when it is able break the UNF-MOD of the m2FE. However, the developed m2FE has guessing gain $1/\Omega$, i.e., $^{\text{UNF-MOD}}\text{Gain}_{\mathcal{A}}(1^{\lambda}) = 1/\Omega$. Hence, we can say that the developed BCMS is unforgeability based secure as the $^{\text{UNF-MOD}}\text{Gain}_{\mathcal{A}}(1^{\lambda})$ of $\mathcal{A}$ is negl ($\lambda$).

**Theorem 3.** *If the underlying hashing technique* SHA256 *is enhanced collision resistance, then $\mathcal{A}$ breaks immutability based security of the proposed BCMS with a negligible probability.*

**Proof.**

**Assumption.** Suppose $\mathcal{A}$ can break the immutability security of the proposed BCMS with a non-negligible advantage.

**Simulator.** We can build a simulator $\mathcal{B}$ to break the security of the underlying SHA256 run by $\mathtt{S}$. $\mathcal{B}$ randomly picks a security parameter $\lambda'$ and provides various oracle queries (e.g., $O_{\text{m2FE.KeyGen}}$, $O_{\text{m2FE.DumGen}}$, $O_{\text{m2FE.Enc}}$, $O_{\text{m2FE.Dec}}$, $O_{\text{m2FE.Ver}}$, $O_{\text{m2FE.Rec}}$, etc.) to $\mathcal{A}$ based on the $\lambda'$, likewise in section IV-*E*.

**Challenge 1.** $\mathcal{A}$ updates the LDS with a forged credential $c_0'$ and deduces $H_{c0}$ (i.e., similar hash of $c_0$) using hash collision property.

**Guess 1.** $\mathcal{A}$ guesses a hash $\Pi$ and sends $\Pi$ to $\mathcal{B}$. Then $\mathcal{B}$ forwards it to $\mathtt{S}$. $\mathcal{A}$ wins if $\mathtt{S}$ finds $H_{c0} = \Pi$. Here, $\mathcal{A}$ is able to find $H_{c0} = \Pi$ only when it is able to find collision in SHA256. However, the SHA256 has a collision rate $1/D$ where $D = 2^{256}$, i.e., $^{\text{IMU-MOD}}{}_1\text{Gain}_{\mathcal{A}}(1^{\lambda}) = 1/2^{256}$. Hence, we can say that the BCMS is immutability based secure as the $^{\text{IMU-MOD}}{}_1\text{Gain}_{\mathcal{A}}(1^{\lambda})$ of $\mathcal{A}$ is negl (D).

**Challenge 2.** $\mathcal{A}$ updates the $n$th block having current block hash $BH_{cur}$ of BCS with a forged credential $c_0'$ and its hash $H_{c0}'$. $\mathcal{A}$ generates the new block hash ($BH_{new}$) similar to $BH_{cur}$ using hash collision property.

**Guess 2.** This guess is similar to **Guess 1**. Hence, we can say that the BCMS is immutability based secure as the $^{\text{IMU-MOD}}{}_2\text{Gain}_{\mathcal{A}}(1^{\lambda}) = \text{negl}(D)$.

## VI. EVALUATION OF BCMS

### A. Experimental Setup

The prototype of the proposed BCMS is developed under the following hardware and software configurations shown in Table II. Besides, it considered 3 INSs (i.e., $INS_1$, $INS_2$, and $INS_3$) and 5 STDs (i.e., $STD_1$, $STD_2$, $STD_3$, $STD_4$, and $STD_5$) holding credential size of 50KB, 100KB, 200KB, 300KB, and 500KB, respectively. Here, we employed RSA encryption with 1024-bit key.

TABLE II: HARDWARE AND SOFTWARE SPECIFICATIONS

| Hardware | Specification | Software | Specification |
|---|---|---|---|
| CPU | 64-bit, 3.10GHz | OS | Windows 11 |
| RAM | 12 GB | IDE | VS Code 2019 |

### B. Experimental Performance

The experimental performances of the proposed BCMS are illustrated in Fig. 6 where Fig. 6(a) presents the registration time required for entity (i.e., STD and INS) by varying the number of entities. It shows that, STD registration requires less time compared to INS registration as STD registration at CMS involves nonce Ñ which is collected from CTA. However, this phase requires almost similar time irrespective to the number of entities because of parallelism service.

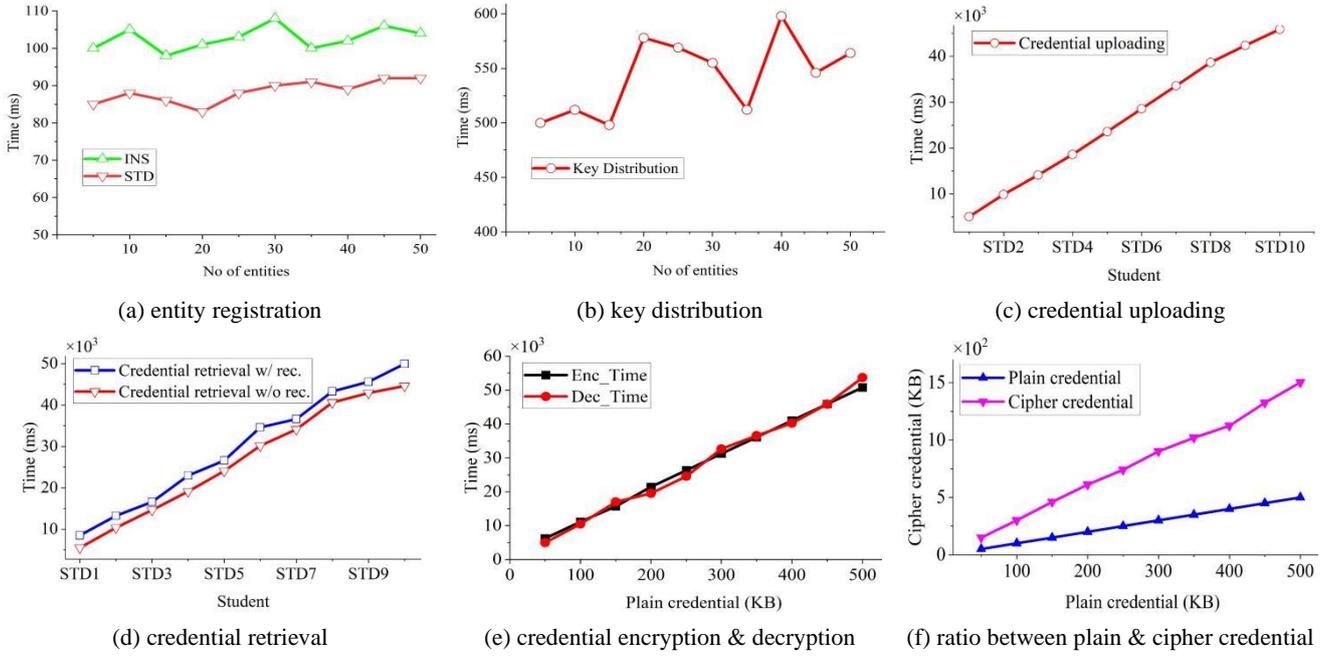

Fig. 6. Experimental Performance.

Fig. 6(b) exhibits the key distribution time required for encryption and decryption of m2FE by varying the number of entities. This phase involving parameter setup and key generation algorithms needs 500ms to 600ms.

Fig. 6(c) shows the time requirement of credential uploading for $STD_1$, ..., $STD_5$ by $INS_1$. It exhibits that the time required for $STD_1$ is less compared to other STDs as $STD_1$ has credential size of 50KB only whereas time required for $STD_5$ is high compared to other STDs as $STD_5$ has credential size of 500KB.

Fig. 6(d) depicts the time requirement of credential retrieval of $STD_1$, ..., $STD_5$ with and without recovery algorithm. It shows that the retrieval time for $STD_1$ with recovery is more compared to retrieval time for $STD_1$ without recovery as it does not need to access BCS and credential updatation in LDS.

Fig. 6(e) presents the time required for encryption and decryption by varying the sizes of plain credential. It shows that time necessity for both encryption and decryption is similar as m2FE involves RSA which has an exponentiation operation both in encryption and decryption. Besides, DNA processing for both algorithms are identical.

Fig. 6(f) demonstrates the ratio between cipher credential and corresponding plain credential. It indicates that the plain and cipher credential ratio is nearly 2.5, i.e., in case of $STD_2$, the cipher credential will be size of around 250KB (as plain credential size is 100KB).

### C. Comparison with other CMSs

#### 1. Theoretical Comparison

Table III presents the theoretical comparison of the proposed BCMS with the other existing systems. It shows that the proposed BCMS focus on student's authentication, credential integrity, privacy, recovery, etc. whereas most of the other CMSs do not concern them. Besides, [18] use very lightweight credential hiding technique and [11] used average credential hiding technique. But in [12], [14], [15], [16] and [17] they don't use any credential hiding technique which is very unsecured.

**Credential Hiding:** In our proposed system, we apply modified two-factor encryption technique for hiding the credential. So, the data are highly secured with reasonable encryption and decryption time. However, as depicted from the Table III, some works hide credential using their distinct

TABLE III. A THEORETICAL COMPARISON BASED ON VARIOUS FEATURES

| Feature | Credential Management System (CMS) | | | | | | | |
|---|---|---|---|---|---|---|---|---|
| | [11] | [12] | [14] | [15] | [16] | [17] | [18] | Proposed |
| Entity authentication by | ID and pass | digital certificate | – | ID and key | OTP | – | ID and pass | email, ID and pass |
| Credential hiding by | 2FE | no | no | symmetric | no | no | asymmetric | m2FE |
| Credential integrity by | BCS | BCS | BCS | BCS | BCS | BCS | BCS | BCS |
| Credential storage by | PV-BCS & LDS | PV-BCS | PB-BCS | PB-BCS | PB-BCS | PB-BCS | PV-BCS | PV-BCS & LDS |
| Credential input by | SA | SA | SA | SA | SA | SA | SA | distributed |
| Credential recovery by | UA | no | no | no | no | no | no | CMS |
| Two factor encryption | yes | no | no | no | no | no | no | yes |

OTP = one-time password; 2FE = two-factor encryption; m2FE = modified 2FE; PV = private; PB = public; BCS = blockchain server; LDS = local dedicated server; SA = single authority; UA = university authority; CMS = credential management system

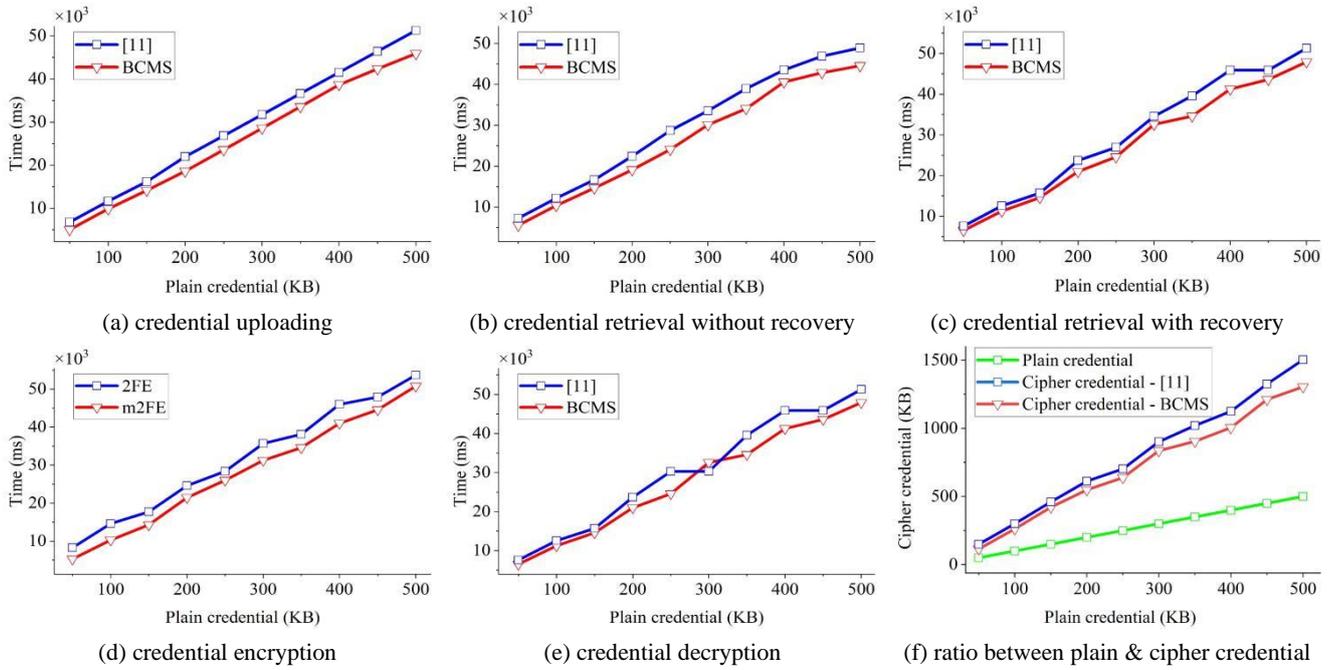
Fig. 7. Comparison with the scheme [11].

ways whereas the others do not consider it.

**Credential integrity:** In our proposed BCMS, we store the credential both in the PV-BCS and LDS that ensures the integrity of the credential. As in Table III, the scheme [11] also achieves the property similar to the proposed one whereas the others employ only either PV-BCS or PB-BCS.

**Modified two-factor encryption:** We apply modified two-factor encryption method which give more secured data and none of the above-mentioned references are use this technique. As in Table III, scheme [11] employed two-factor encryption technique.

**Credent input:** The proposed BCMS stores the credential in a distributed manner whereas all the other schemes gave control to a single user which may become very impractical.

*2. Runtime Comparison*

The developed BCMS is compared with the scheme [11]. To fair comparison, we employed similar setting and the comparisons are illustrated in Fig. 7.

Fig. 7(a) presents the credential uploading phase time of the proposed BCMS and [11] by varying the sizes of plain credential. It shows that the proposed BCMS requires less time compared to [11].

Fig. 7(b) and Fig. 7(c) display the time requirement for the credential retrieval without recovery and with recovery phases in case of both the proposed BCMS and [11] by varying the sizes of plain credential. Both figures indicate that the proposed BCMS requires less time compared to [11].

Fig. 7(d) and Fig. 7(e) show the time necessity for the encryption and decryption of both the m2FE (i.e., developed in proposed BCMS) and 2FE (i.e., used in [11]) by varying the sizes of plain credential. Most of the cases, the m2FE requires less time compared to 2FE.

Fig. 7(f) demonstrates the ratio between cipher credential and corresponding plain credential generated by the both m2FE and 2FE encryption techniques. It indicates that the plain and cipher credential ratio in case of the 2FE is 3 whereas m2FE has 2.5. So, the m2FE produces less cipher size compared to the 2FE.

## VII. CONCLUSION

Credential of a student is crucial and any alteration of it may hamper his social and professional life. Nowadays, almost every university keep credential on local every, which is extremely susceptible to different security threats. In this paper, we propose a distributed, decentralized, and tamper-proof secure system for storing, managing, and verifying credential using the blockchain platform. Here, only authentic users can access the credential and there is negligible chance of credential alteration. We employ a two-factor encryption to enhance credential security. The analysis shows the efficiency of our proposed system. The time required for asymmetric key encryption is too high. So there is a further improvement to use a symmetric key encryption method for reducing the encryption and decryption time by sacrificing a certain privacy level.